\newcommand\be{\begin{equation}}
\newcommand\ee{\end{equation}}
\newcommand\bdm{\begin{displaymath}}
\newcommand\edm{\end{displaymath}}

\documentclass[twocolumn,showpacs,prb]{revtex4}
\usepackage{graphicx}
\usepackage{latexsym}

\begin{document}

\title{Decomposition of the spin-$\frac{1}{2}$ Heisenberg chain compound 
Sr$_{2}$CuO$_{3}$ in air and water:  An EPR and magnetic susceptibility study of
Sr$_{2}$Cu(OH)$_{6}$}
\author{J. M. Hill}
\author{D. C. Johnston}
\author{L. L. Miller}
\affiliation{Ames Laboratory and Department of Physics and Astronomy, Iowa
State University, Ames, Iowa 50011}
\date{\today}
\begin{abstract}\hglue 0.15in
The reaction of Sr$_{2}$CuO$_{3}$ with air and liquid water was studied to address 
the origin of the reported variable Curie-Weiss impurity contribution to the magnetic 
susceptibility $\chi$ of this 
compound at low temperatures. Sr$_{2}$CuO$_{3}$ was found to decompose 
upon exposure to either of these environments. The compound Sr$_{2}$Cu(OH)$_{6}$ was
identified as the primary reaction product.  A pure sample of 
Sr$_{2}$Cu(OH)$_{6}$ was then prepared separately.  Electron 
paramagnetic resonance (EPR), isothermal magnetization versus magnetic 
field $M(H)$ and $\chi$ versus temperature $T$ measurements demonstrate 
that Sr$_{2}$Cu(OH)$_{6}$ contains weakly interacting Cu$^{+2}$ magnetic moments
with spin $S = \frac{1}{2}$ and average $g$-factor 2.133.  From a fit of $\chi(T)$ by the 
Curie-Weiss law and of the $M(H)$ isotherms by modified Brillouin 
functions, the exchange interaction between adjacent Cu$^{+2}$ spins 
was found to be $J/k_{\mathrm{B}} = -1.06(4)$\,K, a weakly 
antiferromagnetic interaction.  Our results indicate that the previously 
reported, strongly sample-dependent, Curie-Weiss contribution to 
$\chi(T)$ of a polycrystalline Sr$_{2}$CuO$_{3}$ sample most likely arises from exposing the sample 
to air, resulting in a variable amount of paramagnetic 
Sr$_{2}$Cu(OH)$_{6}$ on the surface of the sample.
\end {abstract}
\maketitle

\section{Introduction}

The physics of low-dimensional quantum spin systems has been 
intensively studied over the 
past decade due to its relevance to the physics of the 
layered cuprate superconductors and to the variety of unconventional magnetic and
electronic properties exhibited by such materials. The field of 
low-dimensional quantum magnetism has a 
long history dating back to the early 1930s with the publication of 
the Bethe ansatz equations\cite{bethe} from which in 
principle the eigenvalues of the spin $S = \frac{1}{2}$ Heisenberg chain 
can be obtained. By the
early 1990s, research on spin-chain and spin-ladder
materials related to the high-temperature superconductors had become 
a subfield of condensed matter physics. The current experimental work 
on spin-ladders has
been driven by theory but is limited by the lack of known spin-ladder
compounds, particularly metals. Of the cuprates, only 
(Sr,Ca)$_{14}$Cu$_{24}$O$_{41}$ is known to become metallic and 
superconducting, and then only under high pressure.\cite{142441,142441-2} 
However, the interpretation of its properties is complicated by 
the fact that it is comprised of both Cu$_{2}$O$_{3}$ ladder and 
CuO$_{2}$ chain layers. To isolate the physics associated with one or 
the other type of spin configuration, it is desirable to study 
metallic compounds with either chains or ladders, but not both.  For reviews of oxide
spin-ladder and spin-chain compounds see
Refs.~\onlinecite{johnston-chain,johnston-ladd,johnston-vop,dagotto-spin}.  

Sr$_{2}$CuO$_{3}$ is a model spin-$\frac{1}{2}$ linear chain
compound. It has an orthorhombic structure (space group
Immm, Ref.~\onlinecite{teske}) containing Cu$^{+2}$ spins $S = 
\frac{1}{2}$.  The orthorhombic structure is derived from the layered 
tetragonal K$_{2}$NiF$_{4}$ structure by removing lines of oxygen 
atoms parallel to the $b$ axis from within the CuO$_{2}$ layers of the
hypothetical tetragonal K$_{2}$NiF$_{4}$-type 
compound Sr$_{2}$CuO$_{4}$.  Magnetic susceptibility
studies\cite{ami,eggert,motoyama,johnston-sr2cuo3} show this 
compound to be a nearly ideal one-dimensional (1D) spin-$\frac{1}{2}$ Heisenberg
antiferromagnet with a strong intrachain Cu--Cu exchange coupling
$J/k_{\mathrm{B}}$ = 2200 $\pm$ 200\,K, while optical
measurements\cite{suzuura,lorenzana}
yield $J/k_{\mathrm{B}} =$ 2800--3000\,K. On the other hand, theoretical
calculations\cite{drechsler-1} indicate that $J/k_{\mathrm{B}}$ can be no larger
than about 2300\,K in this compound.  Muon spin rotation/relaxation ($\mu$SR) and 
neutron diffraction measurements on single crystals\cite{karen-1,karen-2,kojima} 
revealed long-range antiferromagnetic ordering in this compound with a N\'{e}el temperature
$T_{\mathrm{N}} \simeq$ 5\,K and an 
ordered magnetic moment of $\approx$ 0.06\,$\mu_{\mathrm{B}}/$Cu atom. For the 1D Heisenberg model,
logarithmic terms in the field theory expression for the 
magnetic susceptibility at very low temperatures yield an infinite 
slope as $T$ approaches its finite value at 0\,K.\cite{eggert-chain1,klumper-2,johnston-chain} 
Takigawa {\textit{et~al.}\ }(Refs.~\onlinecite{takigawa-1,takigawa-2,takigawa-3}) 
claim to have seen this behavior in their NMR data:  a 
downturn with decreasing $T$ was observed in the magnetic susceptibility at low $T$, but the 
downturn was not fitted well by the predicted logarithmic behavior. Theory also 
predicts separated spin and charge excitations near the Fermi energy
called ``spinon'' and ``holon'' excitations, respectively, for 1D correlated 
systems (see for example Ref.~\onlinecite{fujisawa}).  Angle-resolved 
photoemission spectroscopy (ARPES) measurements by Fujisawa
{\textit{et~al.}\ }(Ref.~\onlinecite{fujisawa}) along the chains ($b$ axis) show good qualitative
agreement with these theoretical predictions.  They observe two separate
dispersions in the Brillouin zone, one which is reflected about $kb/\pi$ 
(holon) and one which is not (spinon).  However, quantitatively their 
measurements are not fitted well by theory.

A superconducting tetragonal phase, Sr$_{2}$CuO$_{3 + 
\delta}$, has been reported to form under high pressure and to 
exhibit a superconducting transition temperature $T_{\mathrm{c}} 
\approx$ 70\,K.\cite{hiroi-1,han, wangyy,laffez}  However, the samples 
contained low 
superconducting volume fractions and showed semiconducting behavior 
above $T_{\mathrm{c}}$ rather than metallic behavior.  Several 
groups\cite{ami-1,kawashima-1,kawashima-2,lobo} subsequently 
reported high pressure synthesis of nonsuperconducting samples 
and Kawashima {\textit{et~al.}\ }(Ref.~\onlinecite{kawashima-2}) 
suggested that the superconductivity arose from Sr$_{2}$CaCu$_{2}$O$_{y}$ 
impurities.  Tetragonal Sr$_{2}$CuO$_{3 + \delta}$ can also be synthesized at ambient
pressure\cite{bonvalot,mitchell,kato} and those samples were all also nonsuperconducting. 
The available evidence indicates that the oxygen content in this 
compound is variable;\cite{lobo,hiroi-1,mitchell,shimakawa,laffez} $\delta$ 
ranges from 0.08 to 0.9.  Neutron diffraction measurements
carried out on a superconducting and on a nonsuperconducting sample\cite{shimakawa,lobo} found
no major differences between them and could not account for the 
superconductivity.  Both samples showed up to 50\% oxygen vacancies in the CuO$_{2}$
planes as in Sr$_{2}$CuO$_{3}$, rather than in the SrO layers. Transmission electron microscopy 
(TEM) measurements\cite{wangyy,hzhang} confirmed 
the presence of the oxygen vacancies in the CuO$_{2}$ planes.  The 
tetragonal structure of Sr$_{2}$CuO$_{3 + \delta}$ thus evidently arises from a 
random distribution of O vacancies in the CuO$_{2}$ square lattice 
planes, rather than the ordered arrangement of oxygen vacancies in the 
CuO$_{2}$ planes as in orthorhombic Sr$_{2}$CuO$_{3}$.

Due to the very large antiferromagnetic Cu--Cu exchange coupling $J$ in 
Sr$_{2}$CuO$_{3}$, the magnitude of the magnetic susceptibility is 
so low that even small amounts of paramagnetic impurities contribute 
significantly to the observed magnetic susceptibility.  
Polycrystalline samples made by Ami {\textit{et~al.}\ }(Ref.~\onlinecite{ami}) which were
exposed to air showed significant Curie-Weiss contributions, observable most easily at
low temperatures, which obscured the intrinsic spin susceptibility.  The paramagnetic impurity
concentrations in the samples responsible for this behavior were small, equivalent to the 
contribution of 0.4\% spins-$\frac{1}{2}$ (with respect to Cu) with $g$-factor $g =$ 2.
The impurity concentration decreased dramatically to $\approx$ 0.1\% when the samples
were annealed at 600--800\,$^{\circ}$C in nitrogen or at 300--600\,$^{\circ}$C
in low-pressure (6 torr) helium.  It was proposed that paramagnetic oxygen defects due to
the uptake of oxygen from the air may be responsible for the 
Curie-Weiss impurity contribution, but no
test of this proposal was carried out.  Mitchell {\textit{et~al.}\ }and Kato
{\textit{et~al.}\ }(Refs.~\onlinecite{mitchell},~\onlinecite{kato})
synthesized samples of Sr$_{2}$CuO$_{3}$ by dehydration of 
Sr$_{2}$Cu(OH)$_{6}$. Sr$_{2}$Cu(OH)$_{6}$ loses 
two molecules of H$_{2}$O per formula unit upon heating to 400\,$^{\circ}$C in an argon 
atmosphere and forms orthorhombic Sr$_{2}$CuO$_{3}$.  When heated to 
$\sim$ 450\,$^{\circ}$C in oxygen, however, the insulating 
tetragonal form of Sr$_{2}$CuO$_{3 + \delta}$ discussed above is formed.   

In view of the importance of Sr$_{2}$CuO$_{3}$ as a model $S = 
\frac{1}{2}$ antiferromagnetic Heisenberg chain compound, it is 
important to understand the dependence of sample handling on the 
magnetic properties of Sr$_{2}$CuO$_{3}$.
We therefore undertook a study of the chemistry associated with sample handling.  We found 
that Sr$_{2}$CuO$_{3}$ decomposes in air to form Sr$_{2}$Cu(OH)$_{6}$, 
Sr(OH)$_{2}$, Cu(OH)$_{2}$ and SrCO$_{3}$.  Sr$_{2}$Cu(OH)$_{6}$ is 
the main product in this reversible reaction.  Direct exposure of 
Sr$_{2}$CuO$_{3}$ to liquid water results in immediate irreversible decomposition to 
Sr$_{2}$Cu(OH)$_{6}$ which then further decomposes to SrCO$_{3}$ and 
Cu(OH)$_{2}$.  Following Sec.~\ref{sec2} which gives experimental 
details of our work, these chemical reactions will be discussed in 
Sec.~\ref{sec3}. 

In Sec.~\ref{sec4} we discuss the crystallography of 
Sr$_{2}$Cu(OH)$_{6}$ which we synthesized in pure form. 
In Sec.~\ref{sec5} we present and analyze our isothermal magnetization versus magnetic 
field $M(H)$ and magnetic susceptibility $\chi$ versus temperature $T$ 
data for Sr$_{2}$Cu(OH)$_{6}$.  We also report in this section the 
results of room-temperature electron paramagnetic 
resonance (EPR) measurements.  Our results and conclusions are 
summarized in Sec.~\ref{sec6}.  Anticipating that section, we demonstrate that 
Sr$_{2}$Cu(OH)$_{6}$ contains weakly interacting Cu$^{+2}$ magnetic
moments with spin $S = \frac{1}{2}$ and average $g$ factor 2.133.  From a fit of $\chi(T)$ by the 
Curie-Weiss law and the $M(H)$ isotherms by modified Brillouin 
functions, the exchange interaction between adjacent Cu$^{+2}$ spins 
was found to be $J/k_{\mathrm{B}} = -1.06(4)$\,K, a weakly 
antiferromagnetic interaction.

\section{\label{sec2}{Experimental Details}}

Several samples of Sr$_{2}$CuO$_{3}$ were synthesized by calcining stoichiometric 
quantities of 99.995\% pure (metals basis) SrCO$_{3}$ (Aithaca 
Chemical Corp.)\ and CuO (Alfa Aesar) in air at 
950\,$^{\circ}$C for several days, regrinding once per day. A powder x-ray 
diffraction (XRD) pattern taken on a Rigaku x-ray diffractometer with Cu$K\alpha$ radiation
is shown as the top trace in Fig.~\ref{fig1}. A typical
sample showed the reported orthorhombic structure, with lattice 
parameters $a =$ 12.72(4)\,\AA, $b =$ 3.904(8)\,\AA,\ and 
$c =$ 3.496(8)\,\AA\ in good agreement with literature 
values.\cite{teske,ami}  XRD also revealed trace amounts of the SrCO$_{3}$ 
and CuO starting materials in the samples as shown in the top-most x-ray pattern in
Fig.~\ref{fig1}.  

Samples of Sr$_{2}$Cu(OH)$_{6}$ were characterized by XRD
analysis using the above diffractometer.  Samples were mixed with dry KBr
and pelletized for mid-range infrared 
spectroscopy (IR) measurements on a Hartmann and Braun Bomem FT-IR.  
Room-temperature EPR measurements were carried out at 9.5\,GHz on a 
Bruker instrument.  The derivative spectrum, $dI/dB$, shown in Fig.~\ref{fig6}
below was obtained in the usual way as a function of magnetic field, but is 
plotted as a function of the spectroscopic splitting factor ($g$-factor) $g$ to provide direct 
comparison with the spectrum reported in the literature.\cite{friebel,krishnan}
Magnetic susceptibility and magnetization measurements below
300\,K were carried out using a Quantum Design superconducting quantum
interference device (SQUID) magnetometer.  The contribution of ferromagnetic impurities
to the measured magnetization was determined from magnetization 
versus magnetic field isotherms between 75 and 300\,K and was found to be equivalent 
to that of $\sim$ 50 ppm ferromagnetic iron metal impurities; this contribution
is corrected for in Figs.~\ref{fig7}~and~\ref{fig8} below.

\section{\label{sec3}{Decomposition of S\lowercase{r}$_{2}$C\lowercase{u}O$_{3}$}}

We initially suspected that Sr$_{2}$CuO$_{3}$ reacts with air when we observed 
that pristine dark brown Sr$_{2}$CuO$_{3}$ changes color to  blue-gray 
within about thirty minutes of exposure to air.  The 
subsequent XRD pattern contained the expected Sr$_{2}$CuO$_{3}$ peaks, but also 
contained several impurity peaks which could not be identified with remnants of 
the  SrCO$_{3}$ or CuO starting materials.  The above process was repeated with 
additional samples to confirm the results.  We found that the time 
required for the above color change to occur ranged up to several 
days, depending on the relative humidity of the laboratory air, which 
suggested that the samples were reacting with the water vapor in the 
air.  Degraded samples which were heated to 950\,$^{\circ}$C in
air exhibited XRD patterns identical to the XRD pattern of a freshly prepared
Sr$_{2}$CuO$_{3}$ sample (those x-rays were taken with the sample in flowing helium gas to
prevent sample degradation while the x-ray data were accumulated). Therefore we
conclude that the degradation of Sr$_{2}$CuO$_{3}$ in air is reversible.  Although not the
primary focus of this paper, we describe below some preliminary experiments carried out
to investigate the observed sample degradation. 

\begin{figure}
\includegraphics[width=3.5in,keepaspectratio]{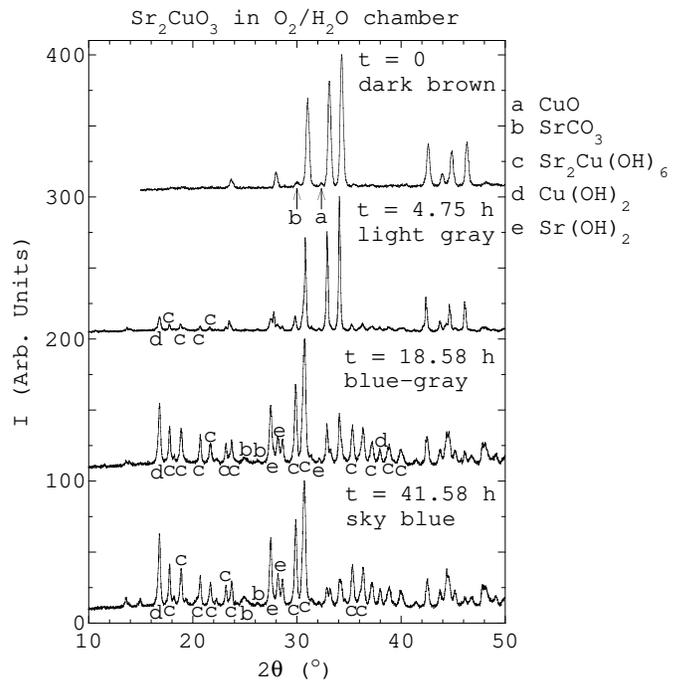}
\vglue 0.1in 
\caption{\label{fig1}{Successive x-ray diffraction patterns 
(Cu$K\alpha$ radiation) showing the decomposition of a Sr$_{2}$CuO$_{3}$ 
sample with time during exposure to flowing hydrated O$_{2}$ gas.  The solid curves are
the diffracted x-ray intensity $I$ versus diffraction angle 2$\theta$ and the letters mark the 
reflections of different impurity phases according to the legend.  The top 
trace is the initial Sr$_{2}$CuO$_{3}$ pattern which shows the presence of small
amounts of the SrCO$_{3}$ and CuO starting materials.
Each x-ray diffraction pattern is scaled so that the most intense reflection has an intensity
of 100.}}
\end{figure}

Since the time scale for sample degradation was clearly humidity 
dependent, for controlled experiments a humidity chamber was constructed 
in which a flow of hydrated 98\% pure nitrogen or oxygen gas was 
passed over a Sr$_{2}$CuO$_{3}$ sample.  The gas was hydrated by diffusing it 
through deionized water.  The relative humidity and temperature 
inside the chamber were measured with a Fisher Scientific Jumbo Thermo-Humidity Meter.  
For sample exposure times up to forty-five hours, the sample 
decomposition results in both gases were identical.  
Figure~\ref{fig1} shows the progression of the x-ray 
diffraction patterns versus time for a Sr$_{2}$CuO$_{3}$ sample 
exposed to hydrated oxygen gas.
The relative humidity of the chamber increased from 50\% to 80\% and 
the temperature ranged from 18.7 to 20.4\,$^{\circ}$C
over the forty-two hour period in Fig.~\ref{fig1}.  The sample decomposed primarily
into Sr$_{2}$Cu(OH)$_{6}$, but small amounts of Cu(OH)$_{2}$, Sr(OH)$_{2}$ and SrCO$_{3}$
could also be identified from XRD patterns as shown in Fig.~\ref{fig1}.
The amount of SrCO$_{3}$ greatly increased when samples were left in the chamber for longer
periods which we attribute to reaction of the sample with the impurity CO$_{2}$ present
in the flowing gas.  

Sr$_{2}$CuO$_{3}$ was next reacted directly with 
deionized liquid water in air and a sky-blue precipitate immediately 
formed. Solutions were stirred for several minutes to ensure 
complete reaction.  During this time the precipitate changed to a mixture 
of black particles and white particles.  The precipitate was allowed to settle and then the
solution was filtered.  XRD analysis of the precipitate showed that it was a
mixture of Cu(OH)$_{2}$ and SrCO$_{3}$.  After heating the mixed precipitate
overnight at 125\,$^{\circ}$C, XRD analysis revealed that the SrCO$_{3}$ was 
unchanged but that the Cu(OH)$_{2}$ had converted to CuO.  The filtrate solution
was kept in a sealed jar for observation.  A substantial amount of white solid appeared
in the solution three to four days later which was identified as 
SrCO$_{3}$ through XRD analysis.  We attribute the formation of SrCO$_{3}$ to the reaction
of unprecipitated Sr$^{+2}$ ions with CO$_{3}^{-2}$ ions and/or dissolved CO$_{2}$ gas present
in the water.  

In order to isolate the primary decomposition product 
Sr$_{2}$Cu(OH)$_{6}$ and minimize formation of SrCO$_{3}$, exposure of the sample to CO$_{2}$ 
must be minimized.  Therefore reaction of a Sr$_{2}$CuO$_{3}$ sample in a 
vacuum-tight vessel with nanopure deionized, degassed water was 
carried out. Two methods of removing gases from the water were used: (i) 
distillation and (ii) repeated sequences of freezing the water from the bottom up in a 
vacuum-sealed glass vessel followed by pumping on the water while melting the ice.  
Initially all samples formed blue or purple-blue precipitates.  The purple
samples may have contained SrCu(OH)$_{4}$ which is a violet-colored sister
compound to Sr$_{2}$Cu(OH)$_{6}$ (see 
Refs.~\onlinecite{scholder},~\onlinecite{ivanov-emin}).
We were not able to confirm the presence of SrCu(OH)$_{4}$ because all of the 
precipitates changed color before they could be isolated.  Samples 
were dried by decanting off as much water as possible, then pumping off 
the residual water.  They were not exposed to the air. All samples 
except one turned color from purple-blue to a shade of green during the drying
process.  The XRD patterns of the green samples (not shown or further 
discussed here) were complex and the phases present in the green 
samples could not be identified.  The purple-blue sample that did not 
change color during the drying process
was identified as primarily Sr$_{2}$Cu(OH)$_{6}$ by XRD analysis. 
The method of degassing the water did not seem to affect the overall 
results of the above experiments which are summarized in 
Table~\ref{table1}. 
\begingroup 
\squeezetable
\begin{table}
\begin{ruledtabular}
\caption{\label{table1}{Summary of reactions of Sr$_{2}$CuO$_{3}$ with 
nanopure deionized, degassed water in a vacuum-tight vessel.  ``Initial 
color'' refers to the color of the solid which immediately formed when 
the Sr$_{2}$CuO$_{3}$ sample contacted the water.  ``Final color'' 
refers to the color of the solid after it had been dried.}}
\begin{tabular}{lll}
METHOD OF DEGASSING WATER & INITIAL COLOR & FINAL COLOR\\ 
freeze/thaw & blue-purple & pale blue\footnote{X-ray had primarily 
Sr$_{2}$Cu(OH)$_{6}$ peaks.}\\
freeze/thaw & pale blue & pale blue-green\\
distilled in N$_{2}$ atmosphere & purple & blue-green\\
distilled in N$_{2}$ atmosphere & dark blue & green\\
distilled in N$_{2}$ atmosphere & sky blue & green\footnote{Turned 
to this color before vacuum pumping began.}\\
\end{tabular}
\end{ruledtabular}
\end{table}
\endgroup 

\section{\label{sec4}{Characterization and crystal structure of S\lowercase{r}$_{2}$C\lowercase{u}(OH)$_{6}$}}

\begin{figure}
\includegraphics[width=3.5in,keepaspectratio]{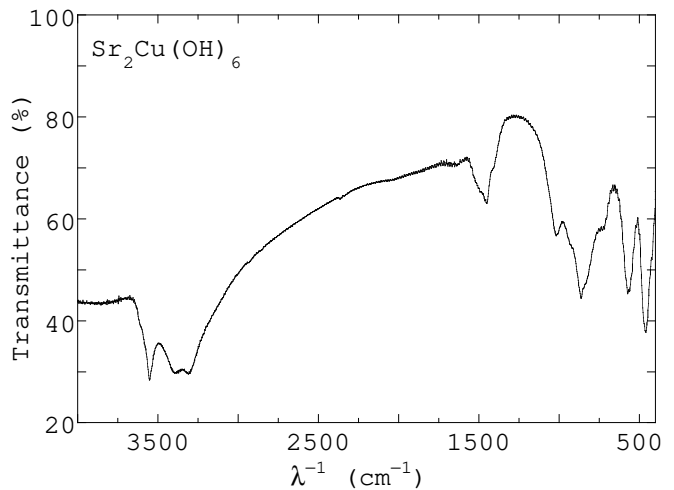}
\vglue 0.1in 
\caption{\label{fig2}{Mid-range infrared spectroscopy spectrum showing 
transmittance versus wavenumber ($\lambda^{-1}$) for Sr$_{2}$Cu(OH)$_{6}$.}}
\end{figure}

In order to characterize the properties of pure Sr$_{2}$Cu(OH)$_{6}$, 
a pure sample of this compound was synthesized in strong hydroxide solution 
following the method of Scholder {\textit{et~al.}\ }(Ref.~\onlinecite{scholder}) using 99.2\% 
Cu(NO$_{3}$)$_{2}\cdot$2$\frac{1}{2}$H$_{2}$O (Fisher Scientific) and 99\%
Sr(OH)$_{2}\cdot$8H$_{2}$O (Alfa Aesar). Figure~\ref{fig2} shows an IR scan 
of the Sr$_{2}$Cu(OH)$_{6}$ sample.  
The scan shows no evidence of the sister compound SrCu(OH)$_{4}$ and 
agrees with literature data.\cite{ivanov-emin}
Inductively coupled plasma (ICP) analysis revealed a Sr:Cu atomic ratio of 2.195 $\pm$ 0.066. 

\begin{figure}
\includegraphics[width=3.5in,keepaspectratio,clip]{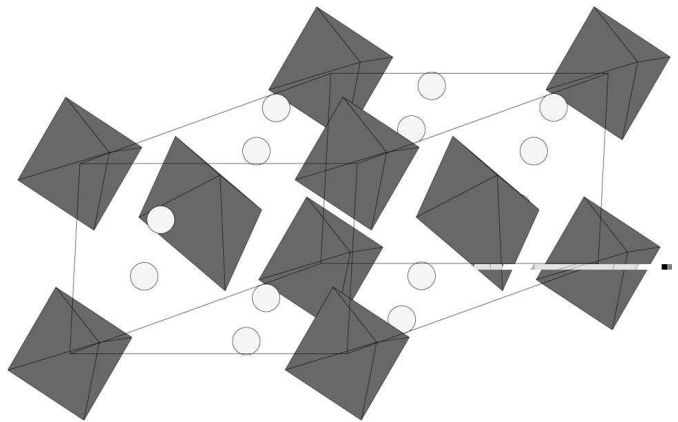}
\vglue 0.1in 
\caption{\label{fig3}{Crystal structure of Sr$_{2}$Cu(OH)$_{6}$.  
The gray octahedra are Cu(OH)$_{6}$ units and the\ spheres 
represent Sr$^{+2}$ ions.}}
\end{figure}

\begin{figure}
\includegraphics[width=3.5in,keepaspectratio]{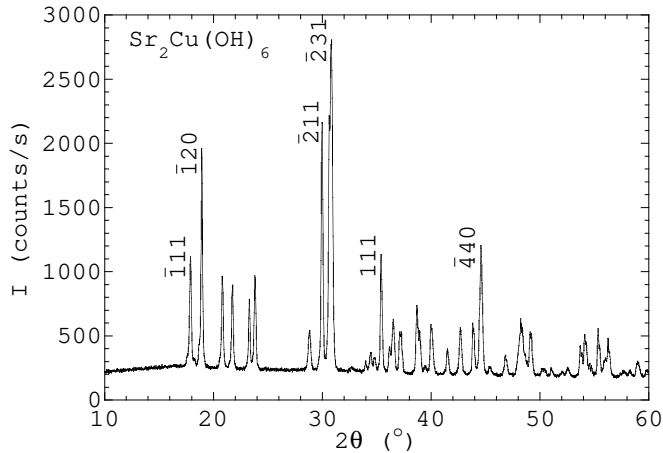}
\vglue 0.1in 
\caption{\label{fig4}{Cu$K\alpha$ x-ray powder diffraction pattern of
Sr$_{2}$Cu(OH)$_{6}$.  The solid curve is intensity $I$ versus 
diffraction angle $2\theta$. The space group is monoclinic P2$_{1}$/b (14) with
$a =$ 8.080(2)\,\AA, $b =$ 9.760(2)\,\AA, $c =$ 6.146(1)\,\AA\ and $\gamma =$
143.64(1)$^{\circ}$. The Miller indices of the six strongest 
reflections are as indicated.}}
\end{figure}

\begin{figure}
\includegraphics[width=3.5in,keepaspectratio]{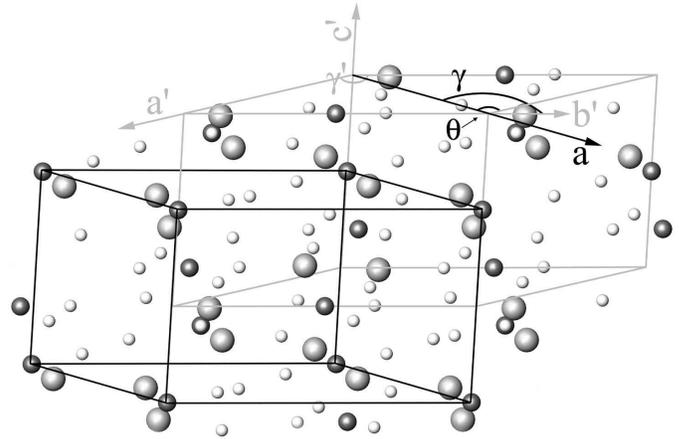}
\vglue 0.1in 
\caption{\label{fig5}{Two alternative unit cells for 
(Ba,Sr)$_{2}$Cu(OH)$_{6}$.  Dubler {\textit{et~al.}\ 
}(Refs.~\onlinecite{dubler-1},~\onlinecite{dubler-2})
used the gray cell in the background with the primed lattice 
parameters for Ba$_{2}$Cu(OH)$_{6}$.  
The black cell in the foreground is an alternate choice and corresponds to the
unit cell used by Nadezhina
{\textit{et~al.}\ }(Refs.~\onlinecite{nadezhina-1},~\onlinecite{nadezhina-2}) for 
Sr$_{2}$Cu(OH)$_{6}$.  $a$ is the short diagonal of the $a'b'$
parallelogram, $b$ and $c$ are equivalent to $b'$ and $c'$, respectively, and
$\gamma$ is the angle between $a$ and $b$.
$\theta$ is the angle between $a$ and $a'$.  Note that the black cell is shifted
in the $c$ direction so that Cu$^{+2}$ ions are on the corners.  Small spheres
represent O$^{-2}$, medium spheres Cu$^{+2}$ and large spheres Ba$^{+2}$ ions.
(Courtesy of Julia~K.~Burzon)}}
\end{figure}

Figure~\ref{fig3} 
shows the crystal structure of Sr$_{2}$Cu(OH)$_{6}$ based on 
structural data from Nadezhina {\textit{et~al.}}\cite{nadezhina-1,nadezhina-2}
This figure emphasizes the highly elongated Jahn-Teller distorted Cu(OH)$_{6}$ octahedra.  
The equatorial Cu-O distances are 1.97 and 1.98\,\AA \ and the apical distance is 
2.63\,\AA. The latter distance is so large that the Cu coordination 
by oxygen should probably be considered to be square planar rather than octahedral.
The Cu(OH)$_{6}$ units are isolated from one another suggesting a weak 
exchange interaction between the Cu$^{+2}$ spins $\frac{1}{2}$.  Figure~\ref{fig4}
shows an x-ray diffraction pattern of a typical sample which we indexed on a monoclinic lattice, 
with space group P2$_{1}$/b (\# 14) and with lattice parameters $a =$ 8.080(2)\,\AA,
$b =$ 9.760(2)\,\AA, $c =$ 6.146(1)\,\AA\ and $\gamma =$ 143.64(1)$^{\circ}$ in agreement
with the results of Nadezhina {\textit{et~al.}}\cite{nadezhina-1,nadezhina-2}
A structure study by Dubler {\textit{et~al.}\ }(Refs.~\onlinecite{dubler-1},~\onlinecite{dubler-2})
reported a different unit cell with different atomic positions in the same space group 
for Ba$_{2}$Cu(OH)$_{6}$.  In order to confirm Dubler {\textit{et~al.}}'s statement that
Ba$_{2}$Cu(OH)$_{6}$ is isostructural to Sr$_{2}$Cu(OH)$_{6}$, we undertook a study of the
relationships of the two respective unit cells and atomic positions after first correcting
for the different space group settings used by the two groups.  Figure~\ref{fig5} shows the 
geometrical relationship between the two unit cells and Table~\ref{table2} lists the
respective lattice parameters.
\begin{table*}
\caption{\label{table2}{Lattice parameters for Ba$_{2}$Cu(OH)$_{6}$ by Dubler 
{\textit{et~al.}\ }(Refs.~\onlinecite{dubler-1},~\onlinecite{dubler-2}) and Sr$_{2}$Cu(OH)$_{6}$ by
Nadezhina {\textit{et~al.}\ }(Refs.~\onlinecite{nadezhina-1},~\onlinecite{nadezhina-2}). The 
Ba$_{2}$Cu(OH)$_{6}$ primed lattice parameters are listed by Dubler in 
a different space group setting. The unprimed lattice parameters correspond 
to the the alternate unit cell used by Nadezhina. The relationship 
between the two unit cells is shown in Fig.~\ref{fig5}.}}
\begin{ruledtabular}
\begin{tabular}{clclcl}
\multicolumn{2}{c}{Ba$_{2}$Cu(OH)$_{6}$ Primed} & \multicolumn{2}{c}
{Ba$_{2}$Cu(OH)$_{6}$ Unprimed} & \multicolumn{2}{c}{Sr$_{2}$Cu(OH)$_{6}$} \\
$a'$ &  6.030(2)\,\AA & $a$ & 8.391(1)\,\AA & $a$ & 8.079(2)\,\AA \\
$b'$ & 10.115(2)\,\AA & $b$ & 10.115(2)\,\AA & $b$ & 9.759(2)\,\AA\\
$c'$ & 6.440(2)\,\AA & $c$ & 6.440(2)\,\AA & $c$ & 6.165(2)\,\AA\\
$\gamma'$ & 124.03(1)$^{\circ}$ & $\gamma$ & 143.44(2)$^{\circ}$ & $\gamma$ & 143.620(1)$^{\circ}$\\ 
vol & 325.5(3)\,\AA$^{3}$ & vol & 325.6(4)\,\AA$^{3}$ & vol & 
288.3(2)\,\AA$^{3}$ \\
\end{tabular}
\end{ruledtabular}
\end{table*}
The two unit cells coincide in the $\hat{z}$ ($c$) direction, but form 
different parallelograms in the $ab$ plane.  The $a$ lattice parameter 
in Nadezhina {\textit{et~al.}}'s unit cell (black cell in the foreground of 
Fig.~\ref{fig5}) is the short diagonal of the parallelogram formed 
by Dubler {\textit{et~al.}}'s 
unit cell (gray cell in the background of Fig.~\ref{fig5}).  
The law of cosines was used to obtain the expressions
\bdm
a = \sqrt{{a'}^{2} + {b'}^{2} + 2a'b'\cos\gamma'}  
\edm
\bdm
b = b'   
\edm
\be
c = c'  
\ee
\bdm
\gamma = 180^{\circ} - \gamma' + \theta    
\edm
\bdm
\theta = \cos^{-1}\left(\frac{{b'}^{2} - a^{2} - {a'}^{2}}{-2aa'}\right) 
\edm
which were used to calculate the unprimed unit cell for Ba$_{2}$Cu(OH)$_{6}$ which 
corresponds to Nadezhina {\textit{et~al.}}'s 
unit cell for Sr$_{2}$Cu(OH)$_{6}$.  The volumes of the unit cells 
are 325.6(4)\,\AA$^{3}$ for Ba$_{2}$Cu(OH)$_{6}$ and 
288.3(2)\,\AA$^{3}$ for Sr$_{2}$Cu(OH)$_{6}$, a difference 
of 37.3(6)\,\AA$^{3}$.  This difference is similar to four times the difference 
between the Ba and Sr atomic 
volumes calculated from structural data for elemental Ba and Sr (Ref.~\onlinecite{pearson}):  
4(62.99\,\AA$^{3}$/atom $-$ 56.325\,\AA$^{3}$/atom) $=$ 
26.66\,\AA$^{3}$/atom  
(the factor of 4 arises because there are two formula units per unit cell). 
Also, since in the same (unprimed) unit cell the $\gamma$ angles of the 
unit cells for the two compounds are essentially the same and the $a$, $b$,
and $c$ lattice parameters for the Ba compound are all $\sim$~4\% larger than
those for the Sr compound, one sees that substituting Ba for Sr results in a
uniform increase in unit cell size.

The fractional atomic positions in the primed unit cell for Ba$_{2}$Cu(OH)$_{6}$ can be 
expressed in terms of the unprimed unit cell according to
\be
\label{eq2}
\left( \begin{array}{c}
x/a \\ y/b \\ z/c
\end{array} \right)
=
- \left[ \left( \begin{array}{ccc}
-\frac{a'\sin\gamma'}{a\sin\gamma} & 0 & 0 \\
\frac{a'\sin(\gamma' + \gamma)}{b\sin\gamma} & \frac{b'}{b} & 0 \\
0 & 0 & \frac{c'}{c}
\end{array} \right)
\left( \begin{array}{c}
x'/a' \\ y'/b' \\ z'/c'
\end{array} \right)
-
\left( \begin{array}{c}
0 \\ 0 \\ \frac{1}{2}
\end{array} \right) \right]  .
\ee
The results are shown in Table~\ref{table3}.
\begingroup
\squeezetable
\begin{table*}
\caption{\label{table3}{Atomic positions for Ba$_{2}$Cu(OH)$_{6}$ by Dubler 
{\textit{et~al.}\ }(Refs.~\onlinecite{dubler-1},~\onlinecite{dubler-2}, primed unit 
cell) and Sr$_{2}$Cu(OH)$_{6}$ by Nadezhina {\textit{et~al.}\ 
}(Refs.~\onlinecite{nadezhina-1},~\onlinecite{nadezhina-2}, unprimed unit cell). The 
`primed' atomic positions for Ba$_{2}$Cu(OH)$_{6}$ correspond to the 
`primed' unit cell in Table~\ref{table2}.  The `unprimed' atomic positions for
Ba$_{2}$Cu(OH)$_{6}$ are obtained by expressing the primed positions in terms of the 
unprimed unit cell listed in Table~\ref{table2} [see Eq.~(\ref{eq2})].  These
unprimed positions are similar to those obtained by Nadezhina {\textit{et~al.}\ }for
Sr$_{2}$Cu(OH)$_{6}$.}}
\begin{ruledtabular}
\begin{tabular}{llllllllll}
  & \multicolumn{3}{c}{Ba$_{2}$Cu(OH)$_{6}$ Primed} & 
\multicolumn{3}{c}{Ba$_{2}$Cu(OH)$_{6}$ Unprimed}  & \multicolumn{3}{c}
{Sr$_{2}$Cu(OH)$_{6}$}\\ \cline{2-4}\cline{5-7}\cline{8-10}
 & \multicolumn{1}{c}{$x'/a'$} & \multicolumn{1}{c}{$y'/b'$} & \multicolumn{1}
{c}{$z'/c'$} & \multicolumn{1}{c}{$x/a$} & \multicolumn{1}{c}{$y/b$} & 
\multicolumn{1}{c}{$z/c$} & \multicolumn{1}{c}{$x/a$} & \multicolumn{1}{c}{$y/b$} & 
\multicolumn{1}{c}{$z/c$} \\
Ba, Sr & 0.2821(1) & 0.0674(1) & 0.2489(1) & 0.2820(4) & 0.0332(5) & 0.4326(1) & 0.2866(2) & 0.0367(2) & 
0.4256(2) \\
Cu & 0 & $\frac{1}{2}$ & 0 & 0 & $\frac{1}{2}$ & $\frac{1}{2}$ &  0 & $\frac{1}{2}$ & $\frac{1}{2}$ \\
O1 & 0.4327(8) & 0.2575(6) & 0.0586(4) & 0.4326(13) & 0.3740(17) & 0.2425(8)  &
0.429(2) & 0.366(2) & 0.233(2) \\
O2 & 0.1956(8) & 0.7629(5) & 0.0393(4) & 0.1956(10) & 0.1563(14) & 0.7371(10)
& 0.202(2) & 0.156(2) & 0.724(2) \\
O3 & 0.1387(8) & 0.4750(6) & 0.2267(4) & 0.1387(9) & 0.9120(14) & 0.0250(9) &
0.122(2) & 0.889(2) & 0.033(2) \\
\end{tabular}
\end{ruledtabular}
\end{table*}
\endgroup
Although the unprimed atomic positions for Ba$_{2}$Cu(OH)$_{6}$ 
do not match those of Sr$_{2}$Cu(OH)$_{6}$ within the errors, the 
close similarities of the respective values demonstrate that Sr$_{2}$Cu(OH)$_{6}$ and 
Ba$_{2}$Cu(OH)$_{6}$ are isostructural.  
Therefore, the primed unit cell used by Dubler {\textit{et~al.}\ }is an
alternative unit cell for the two compounds.

\section{\label{sec5}{EPR, Magnetic Susceptibility, and Magnetization of S\lowercase{r}$_{2}$C\lowercase{u}(OH)$_{6}$}}

\begin{figure}
\includegraphics[width=3.5in,keepaspectratio]{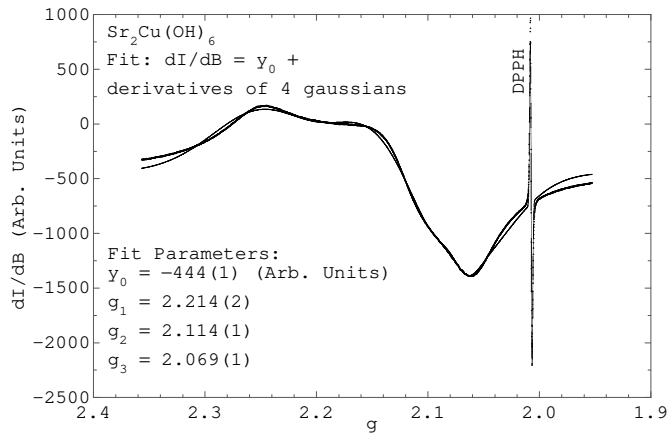}
\vglue 0.1in 
\caption{\label{fig6}{The thick curve is a powder EPR derivative spectrum,
$dI/dB$ versus spectroscopic splitting factor $g$ of Sr$_{2}$Cu(OH)$_{6}$ at room
temperature using an rf frequency of 9.5\,GHz (X-band).  The thin curve 
is a multiple gaussian derivative fit to the data with parameters shown in the figure.
DPPH ($g =$ 2.0036) was used as an internal magnetic field standard.}}
\end{figure}

Figure~\ref{fig6} shows a typical room-temperature EPR spectrum of a powder
Sr$_{2}$Cu(OH)$_{6}$ sample and a DPPH internal standard.
The hyperfine interaction of the Cu$^{+2}$ electronic spin-$\frac{1}{2}$ with the
Cu nuclear spin $I = \frac{3}{2}$ has a typical width 
of 20--100\,G,\cite{orton-1} but it is not resolved in our data.  We 
believe this is due to several factors.   We expect to see ``absorption-like''
features rather than sharp derivative peaks since the material is a powder.\cite{pilbrow}  
At room temperature, spin-lattice relaxation leads to broadened features 
which obscure the hyperfine peaks.\cite{bowers,abrablea} Since our 
system is not magnetically dilute, the spin-spin 
interaction also leads to peak broadening.\cite{abrablea} 

The function used to fit the EPR data consisted of a vertical offset term and
the sum of the derivatives of four gaussians (including one for the 
DPPH magnetic field marker) which yielded three principal-axis $g$ values for 
Sr$_{2}$Cu(OH)$_{6}$ consistent with the rhombic symmetry of the Cu 
site.  The DPPH-corrected $g$ values, 2.214(2), 2.114(1), and 2.069(1), are in 
agreement with the literature values.\cite{friebel,krishnan} In order 
to incorporate these experimentally determined values into fits to the powder
magnetic susceptibility and magnetization data, the spherical (powder) average 
must be used. The Curie constant which occurs in the magnetic susceptibility fit function
[Eqs.~(\ref{eq5},~\ref{eq6},~\ref{eq8}) below] is a function of 
$g^{2}$; therefore, the appropriate average of $g$ is the rms $g$ value, $g_{\mathrm{A}}$,
as given in Eq.~(\ref{eq3}).
The Brillouin function [Eq.~(\ref{eq10}) below] used to fit our low temperature
magnetization data is a function of the average of $g$ itself, as given 
by $g_{\mathrm{B}}$ in Eq.~(\ref{eq4}).
Not surprisingly, these two values are nearly identical.
\be
\label{eq3}
g_{\mathrm{A}} = \sqrt{\frac{\left(g_{1}^{2} + g_{2}^{2} + 
g_{3}^{2}\right)}{3}} = 2.133
\ee
\be
\label{eq4}
g_{\mathrm{B}} = \frac{\left(g_{1} + g_{2} + g_{3}\right)}{3} = 2.132
\ee

\begin{figure}
\includegraphics[width=3.5in,keepaspectratio]{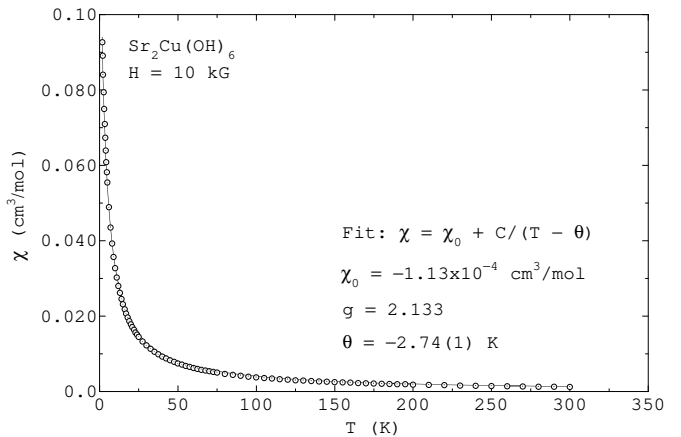}
\vglue 0.1in 
\caption{\label{fig7}{Magnetic susceptibility $\chi$ versus 
temperature $T$ of 
Sr$_{2}$Cu(OH)$_{6}$ ($\odot$). The solid curve is a fit to the data by the
function shown in the figure [Eq.~(\ref{eq5})], with parameters also listed
in the figure where $g$ is $g_{\mathrm{A}}$ as given in Eq.~(\ref{eq3}).}}
\end{figure}

The magnetic susceptibility $\chi$ versus temperature $T$ in an applied 
magnetic field $H =$ 10\,kG is shown in Fig.~\ref{fig7}.  
We fitted the data by
\be
\label{eq5}
\chi = \chi_{0} + \frac{C}{T - \theta}  \ ,
\ee
where $\theta$ is the Weiss temperature and $C$ is the Curie constant given by
\be
\label{eq6}
C = \frac{Ng^{2}\mu_{\mathrm{B}}^{2}S\left(S + 1\right)}{3k_{\mathrm{B}}}
\ee
in which $N$ is the number of spins in the sample, 
$g$ is $g_{\mathrm{A}}$ (Eq.~(\ref{eq3}), $\mu_{\mathrm{B}}$ is the Bohr magneton, $S$ is the
spin of the Cu$^{+2}$ ion (assumed to be $\frac{1}{2}$) and
$k_{\mathrm{B}}$ is the Boltzmann constant. In the molar units of 
$\chi$ or $M$ in 
Figs.~\ref{fig7},~\ref{fig8},~\ref{fig10},~and~\ref{fig11} below,
$N$ is set to $N_{\mathrm{A}}$ (Avogadro's number).
The $T$-independent $\chi_{0}$ term 
\be
\chi_{0} = \chi^{\mathrm{core}} + \chi^{\mathrm{VV}}
\ee
is the sum of the contribution from the diamagnetic cores of the 
atoms $\chi^{\mathrm{core}}$ plus the paramagnetic Van Vleck susceptibility $\chi^{\mathrm{VV}}$
of the Cu$^{+2}$ ions.  

A fit to all the $\chi(T)$ data in Fig.~\ref{fig7} by 
Eq.~(\ref{eq5}) with $\chi_{0}$ set to the diamagnetic core contribution for 
Sr$_{2}$Cu(OH)$_{6}$ ($-1.13\times10^{-4}$\,cm$^{3}/$mol), 
yields the fit (solid curve) in Fig.~\ref{fig7} with a Weiss temperature
$\theta = -$2.74(1)\,K indicating weak coupling between the Cu$^{+2}$ 
spins-$\frac{1}{2}$,
as expected.  The negative sign of $\theta$ corresponds to an 
antiferromagnetic coupling between the Cu spins.  When 
$\chi_{0}$ was allowed to vary, $\chi_{0}$ became more negative than the diamagnetic core
contribution, which is physically unreasonable.  We were able to obtain a better fit
\begin{figure}
\includegraphics[width=3.5in,keepaspectratio]{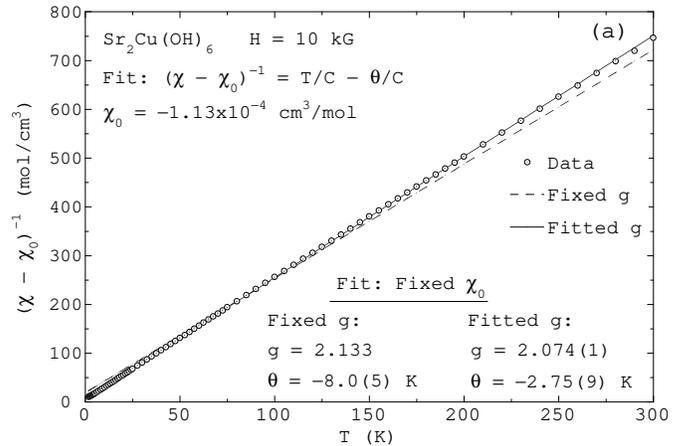}
\vglue 0.1in 
\includegraphics[width=3.5in,keepaspectratio]{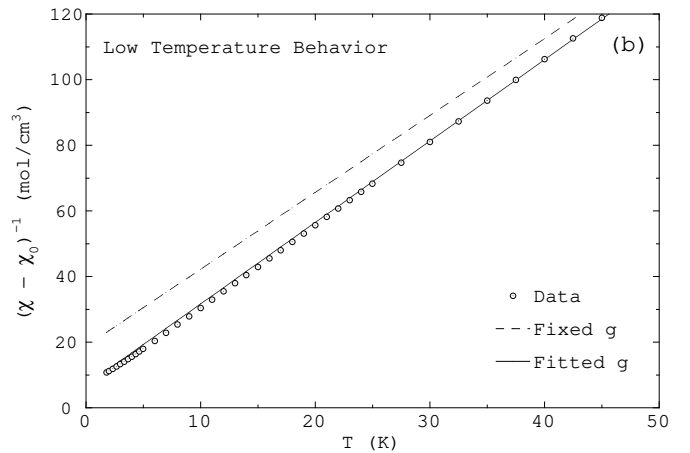}
\vglue 0.1in 
\caption{\label{fig8}{(a) Inverse magnetic susceptibility corrected for the contribution 
of $\chi_{0}$, $(\chi~-~\chi_{0})^{-1}$, versus temperature $T$
($\odot$) of Sr$_{2}$Cu(OH)$_{6}$. The dashed line is the ``Fixed g'' fit yielding the 
$\theta$ parameter shown in the figure where the fixed 
$g$ is $g_{\mathrm{A}}$ in Eq.~(\ref{eq3}). The solid line is the
``Fitted g'' fit which yields the indicated $g$ and $\theta$ values. $\chi_{0}$ is 
fixed at $\chi^{\mathrm{core}}$ for both fits.  (b) Expanded plot of 
the low temperature data and fits below 50\,K.}}
\end{figure}
when $C$ was allowed to vary.  However, the fitted $C$ value yielded 
a $g$ value from Eq.~(\ref{eq6}) which was significantly lower than the measured
average $g$ value obtained from EPR. 

Figure~\ref{fig8} shows the inverse of the magnetic susceptibility corrected for the contribution of
$\chi_{0}$, $(\chi~-~\chi_{0})^{-1}$, versus temperature $T$ in an applied magnetic field
$H =$ 10\,kG.  The dashed line is a linear fit 
\be
\label{eq8}
\frac{1}{\chi - \chi_{0}} = \frac{T - \theta}{C}
\ee
[see Eq.~(\ref{eq5})] with fixed $C$ given by Eq.~(\ref{eq6}) which yields
$\theta = -$8.0(5)\,K.  This $\theta$ is significantly larger in magnitude 
than obtained from the $\chi(T)$ fit in Fig.~\ref{fig7}.
The solid line in Fig.~\ref{fig8} is a linear fit with fitted $C$ and is clearly a better fit to
the data.  Although the latter $\theta = -$2.75(9)\,K agrees with 
that from the fit in Fig.~\ref{fig7}, the average $g =$ 2.074(1) obtained from $C$
is lower than the average value obtained from EPR.  We could not
obtain an optimum fit to our data with physically reasonable parameters 
using the $g$ value from the EPR measurements.  At low temperatures, shown 
in Fig.~\ref{fig8}(b),
both the ``Fitted g'' and the ``Fixed g'' fits deviate from the data.  
  
As noted above, the $\theta$ values obtained from the fixed-$g$ fits to $\chi(T)$ and 
$(\chi~-~\chi_{0})^{-1}$ versus $T$ do not agree.   Fitting $\chi(T)$ emphasizes the  
low-temperature regime where $\chi$ is varying most strongly with $T$ 
due to the Curie-Weiss behavior.  The $(\chi~-~\chi_{0})^{-1}$ data, however, 
emphasize the high-temperature behavior, where weak temperature 
dependence of $\chi_{0}$ and/or the contribution to $\chi(T)$ from 
small amounts of impurities could most strongly influence the 
parameters obtained from the fit.  Therefore, the parameters 
obtained from the one-parameter $\chi(T)$ fit,
\bdm
\chi_{0} = -1.13\times10^{-4}\,{\mathrm{cm}}^{3}/{\mathrm{mol}}
\edm
\be
\label{eq9}
g = 2.133
\ee
\bdm
\theta = -2.74(1)\,{\mathrm{K}}  \ ,
\edm
are considered to be more reliable and best represent the intrinsic 
behavior of Sr$_{2}$Cu(OH)$_{6}$.  

\begin{figure}
\includegraphics[width=3.5in,keepaspectratio]{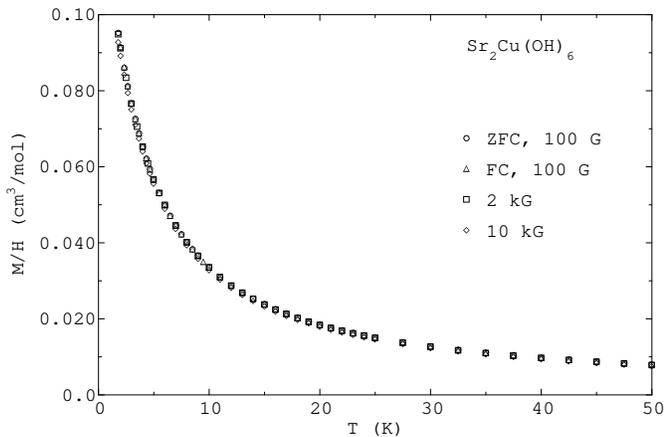}
\vglue 0.1in 
\caption{\label{fig9}{Magnetization divided by magnetic field $M/H$ 
versus temperature $T$ for Sr$_{2}$Cu(OH)$_{6}$. The zero-field-cooled (ZFC) and
field-cooled (FC) data ($\circ$ and $\triangle$, respectively) were taken in an
applied magnetic field of $H =$ 100\,G.  Also shown are data taken in 
$H =$ 2\,kG ($\Box$) and 10\,kG ($\Diamond$).}}
\end{figure}

To investigate the low temperature behavior further, several 
magnetization versus applied magnetic field $M(H)$ isotherms at low temperatures and both 
zero-field-cooled (ZFC) and field-cooled (FC) $M(T)$ data at 
$H =$ 100\,G were taken.  The ZFC and FC data show no evidence of long-range ordering
above 1.8\,K as shown in Fig.~\ref{fig9}.  
The $M(H)$ isotherms at low temperatures are shown in Fig.~\ref{fig10}.  
\begin{figure}
\includegraphics[width=3.5in,keepaspectratio]{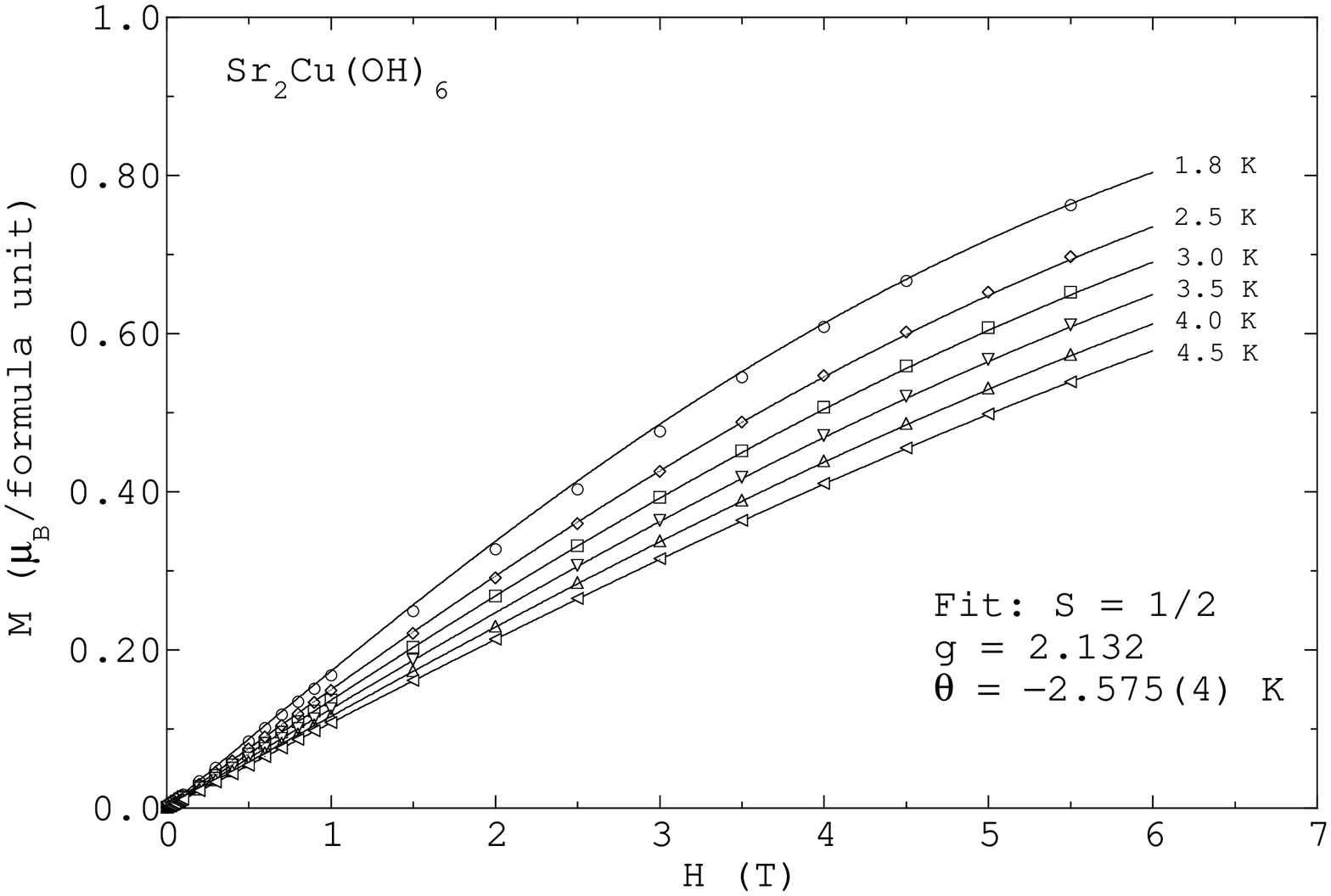}
\vglue 0.1in 
\caption{\label{fig10}{Magnetization $M$ versus field $H$ for 
Sr$_{2}$Cu(OH)$_{6}$ at 1.8\,K ($\circ$), 2.5\,K ($\Diamond$), 3.0\,K 
($\Box$), 3.5\,K ($\bigtriangledown$), 4.0\,K ($\triangle$) and 
4.5\,K ($\lhd$). The solid curves are a fit to the data using 
Eq.~(\ref{eq10}) with parameters shown in the figure where 
$g$ is $g_{\mathrm{B}}$ as given in Eq.~(\ref{eq4}).}}
\end{figure}
The data up to $H =$ 1\,T are in the low-field proportional 
part of the $M(H)$ curves, which explains why all the magnetization data 
in Fig.~\ref{fig9} lie on a common curve.  

We obtained a robust fit to the $M(H)$ isotherm data in 
Fig.~\ref{fig10} using a modified Brillouin function\cite{kittel} for $S = \frac{1}{2}$
\be
\label{eq10}
M = NgS\tanh\left[\frac{gS\mu_{\mathrm{B}}H}{k_{\mathrm{B}}(T - \theta)}\right] \ ,
\ee
where $g$ is $g_{\mathrm{B}}$ as given in Eq.~(\ref{eq4}), and $T$ in the usual 
Brillouin function\cite{kittel} is replaced by $T - \theta$.  This change was 
necessary so that the high-temperature and/or low-field expansion of 
Eq.~(\ref{eq10}) yielded the observed Curie-Weiss behavior $M = 
CH/(T~-~\theta)$.  The fit yielded $\theta = -$2.575(4)\,K. This
value for $\theta$ agrees with the value in Eq.~(\ref{eq9})
obtained from the fit to the magnetic susceptibility data, as it should.
A comparison of the two values gives the estimate $\theta = -2.66(9)$\,K.
When we allowed the spin $S$ to vary during a fit, the fitted $S$ value ranged from 0.471 
to 0.516 indicating that the spin is indeed $\frac{1}{2}$ as expected for Cu$^{+2}$.
Allowing $g$ to vary at fixed $S = \frac{1}{2}$ produced a slightly better fit,
but with an incorrect $g$ value ($g =$ 2.179 compared to the actual value 2.132).
In Fig.~\ref{fig11}, 
\begin{figure}
\includegraphics[width=3.5in,keepaspectratio]{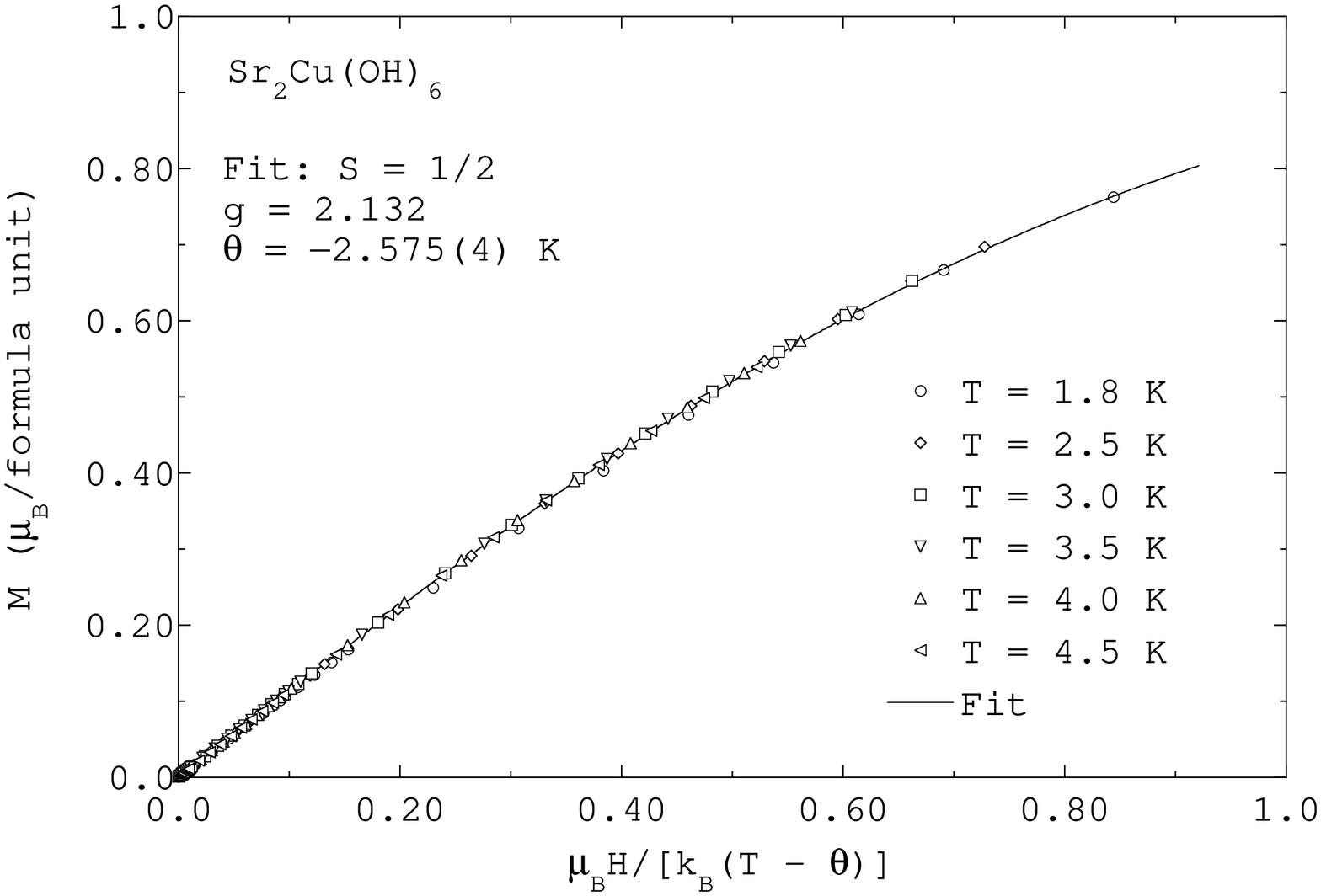}
\vglue 0.1in 
\caption{\label{fig11}{Magnetization $M$ versus the ratio 
of magnetic field energy to the modified thermal energy 
$\mu_{\mathrm{B}}H/k_{\mathrm{B}}(T~-~\theta)$ for 
Sr$_{2}$Cu(OH)$_{6}$ at 1.8\,K ($\circ$), 2.5\,K ($\Diamond$), 3.0\,K 
($\Box$), 3.5\,K ($\bigtriangledown$), 4.0\,K ($\triangle$) and 
4.5\,K ($\lhd$). The solid curve is a fit to all the data by 
Eq.~(\ref{eq10}), with fitting parameter $\theta$, fixed $g = g_{\mathrm{B}}$
as given in Eq.~(\ref{eq4}) and $S =$ 1/2.}}
\end{figure}
a scaling plot of the magnetization $M$ versus the ratio of magnetic field energy to the
modified thermal energy $\mu_{\mathrm{B}}H/k_{\mathrm{B}}(T~-~\theta)$ is 
shown and we see that the fit (solid curve) does indeed reproduce the data very well. 

In summary, we find that the best description of the combined EPR, 
$\chi(T)$ and $M(H)$ data for Sr$_{2}$Cu(OH)$_{6}$ is that the 
Cu$^{+2}$ ions have spin $S = \frac{1}{2}$ with $g =$ 2.133; the Weiss temperature
in the Curie-Weiss law is $\theta = -2.66(9)$\,K.  
Assuming a Heisenberg interaction between nearest-neighbor spins with Hamiltonian 
$\mathcal{H}~=~-J{\displaystyle{\sum_{<i~j>}}}~{\vec{S}}_{i}~\cdot~{\vec{S}}_{j}$, 
where the sum is over all distinct nearest-neighbor pairs of spins and $J~>$~0 ($J~<$~0) 
corresponds to a ferromagnetic (antiferromagnetic) interaction, the exchange constant
$J$ is given in terms of $\theta$ by $J~=~3k_{\mathrm{B}}\theta/[zS(S~+~1)]$ where
$z$ is the number of nearest neighbors.\cite{kittel}  In 
Sr$_{2}$Cu(OH)$_{6}$, each Cu atom has 10 Cu nearest neighbors ($z =$ 
10) at a distance of 5.8--6.2\,\AA; the Cu next-nearest neighbors are 
at distances of $\geq 8.1$\,\AA.  Using $\theta = -2.66(9)$\,K, one 
thus obtains $J/k_{\mathrm{B}} = -1.06(4)$\,K.

\section{\label{sec6}{Summary and Conclusions}}

We have demonstrated that Sr$_{2}$CuO$_{3}$ decomposes in
both air and liquid water and that the primary decomposition product 
is Sr$_{2}$Cu(OH)$_{6}$. In contrast, the compound La$_{2}$CuO$_{4}$ 
can be successfully electrochemically oxidized in aqueous base without
any noticeable decomposition.\cite{chou-4} 

The magnetic susceptibility of Sr$_{2}$Cu(OH)$_{6}$ exhibits Curie-Weiss
behavior down to low temperatures and indicates only very weak interactions between
the Cu$^{+2}$ spins.  The crystallography, EPR, and magnetization measurements are
consistent with a nearly isolated, spin $S = \frac{1}{2}$, local moment model for
Sr$_{2}$Cu(OH)$_{6}$.  We obtained unusually good consistency between the $M(H)$ and $\chi(T)$ 
fits which yielded a small $\theta = -2.66(9)$\,K.  The spherically 
averaged $g$ of the Cu$^{+2}$ spins is 2.133 obtained from EPR
and is similar to those of other cuprates.  For example, $g$ in CuO is 2.125(5); in 
La$_{2}$BaCuO$_{5}$ and in powder Sr$_{14}$Cu$_{24}$O$_{41}$ it is 2.103 
and 2.14, respectively (from Table \textrm{V} in Ref.~\onlinecite{johnston-ladd}). 

Since the magnitude of the magnetic susceptibility of the linear chain 
compound Sr$_{2}$CuO$_{3}$ is small due to the strong antiferromagnetic coupling between
the Cu spins, one would expect even a small impurity concentration of Sr$_{2}$Cu(OH)$_{6}$
to produce a significant paramagnetic contribution at low temperatures.  Although we
cannot rule out the possibility that paramagnetic oxygen species are generated
upon exposure of Sr$_{2}$CuO$_{3}$ to air as proposed by Ami 
{\textit{et~al.}\ }(Ref.~\onlinecite{ami}),
our experiments indicate that the reported variable Curie-Weiss contributions to the
magnetic susceptibility of polycrystalline Sr$_{2}$CuO$_{3}$ were most 
likely mainly due to varying amounts of Sr$_{2}$Cu(OH)$_{6}$ on the 
sample surfaces due to exposure of the sample to the humidity in 
the air.  

The Cu--Cu exchange coupling $J/k_{\mathrm{B}} = -1.06(4)$\,K in 
Sr$_{2}$Cu(OH)$_{6}$ is very weak compared to $J/k_{\mathrm{B}} \sim -1600$\,K
in the high-$T_{\mathrm{c}}$ cuprate superconductors, due to the isolated
square-planar coordination of the Cu$^{+2}$ ions in Sr$_{2}$Cu(OH)$_{6}$.  The nearest-neighbor
Cu--Cu exchange path is Cu--O--O--Cu with a zig-zag geometry and a 
Cu--Cu distance of 5.8\,\AA, whereas in the planar high-$T_{\mathrm{c}}$ 
cuprates the nearest-neighbor distance is 2.80\,\AA\ with a strong 
180\,$^{\circ}$ Cu--O--Cu antiferromagnetic superexchange coupling.
Thus Sr$_{2}$Cu(OH)$_{6}$ serves as nice reference material for comparison
to the magnetic properties of more strongly interacting systems such as the
high-$T_{\mathrm{c}}$ cuprates.

\begin{acknowledgments}
We thank Joseph Shinar for useful discussions regarding the EPR spectrum and Paul
K\"{o}gerler for his help with the IR and EPR measurements.
Ames Laboratory is operated for the U.S. Department of Energy by Iowa State
University under Contract No.\ W-7405-Eng-82.  This work was supported by 
the Director for Energy Research, Office of Basic Energy Sciences.
\end{acknowledgments}

\end{document}